\documentclass[10pt,conference]{IEEEtran}
\usepackage{cite}
\usepackage{amsmath,amssymb,amsfonts}
\usepackage{algorithmic}
\usepackage{graphicx}
\usepackage{textcomp}
\usepackage{xcolor}
\usepackage[hyphens]{url}
\usepackage{fancyhdr}
\usepackage{balance}

\usepackage[bookmarks=true,breaklinks=true,letterpaper=true,colorlinks,citecolor=blue,linkcolor=blue,urlcolor=blue]{hyperref}

\newcommand{\hpcayear}{2026}

\usepackage{physics}
\usepackage{algorithmic}
\usepackage{multirow}
\usepackage{amsthm}
\usepackage{enumitem}
\usepackage{tikz}
\usepackage[ruled]{algorithm2e}

\newcommand{\un}[1]{\underline{#1}}

\newcommand{\ignore}[1]{}
\newcommand{\revignore}[1]{}

\definecolor{OliveGreen}{HTML}{1A4D2E}
\definecolor{orange}{HTML}{1A4D2E}

\definecolor{aliceblue}{rgb}{0.94, 0.97, 1.0}
\usepackage{xcolor}
\usepackage{tcolorbox}
\tcbset{width=1\columnwidth,boxrule=1pt,colback=aliceblue,arc=1pt,left=2pt,right=2pt,boxsep=0.5pt}

\newtcolorbox{hintbox}[2][]
{
  colframe = oldaliceblue!100,
  colback  = oldaliceblue!5,
  boxsep=2pt,
  width=\dimexpr\columnwidth\relax, 
  coltitle = oldaliceblue!20!black,
  title    = #2,
  #1,
}

\newcommand{\hpcasubmissionnumber}{868}
\title{BARD: Reducing Write Latency of DDR5 Memory\\by Exploiting Bank-Parallelism}

\def\hpcacameraready{} 

\newcommand\hpcaauthors{Suhas Vittal and Moinuddin Qureshi}
\newcommand\hpcaaffiliation{Georgia Institute of Technology}
\newcommand\hpcaemail{\{suhaskvittal, moin\}@gatech.edu}

\author{
  \ifdefined\hpcacameraready
    \IEEEauthorblockN{\hpcaauthors{}}
      \IEEEauthorblockA{
        \hpcaaffiliation{} \\
        \hpcaemail{}
      }
  \else
    \IEEEauthorblockN{\normalsize{HPCA \hpcayear{} Submission
      \textbf{\#\hpcasubmissionnumber{}}} \\
      \IEEEauthorblockA{
        Confidential Draft \\
        Do NOT Distribute!!
      }
    }
  \fi 
}

\fancypagestyle{camerareadyfirstpage}{%
  \fancyhead{}
  
  \fancyhead[C]{
    \ifdefined\aeopen
    \parbox[][12mm][t]{13.5cm}{\hpcayear{} IEEE International Symposium on High-Performance Computer Architecture (HPCA)}    
    \else
      \ifdefined\aereviewed
      \parbox[][12mm][t]{13.5cm}{\hpcayear{} IEEE International Symposium on High-Performance Computer Architecture (HPCA)}
      \else
      \ifdefined\aereproduced
      \parbox[][12mm][t]{13.5cm}{\hpcayear{} IEEE International Symposium on High-Performance Computer Architecture (HPCA)}
      \else
      \parbox[][0mm][t]{13.5cm}{\hpcayear{} IEEE International Symposium on High-Performance Computer Architecture (HPCA)}
    \fi 
    \fi 
    \fi 
    \ifdefined\aeopen 
      \includegraphics[width=12mm,height=12mm]{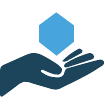}
    \fi 
    \ifdefined\aereviewed
      \includegraphics[width=12mm,height=12mm]{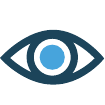}
    \fi 
    \ifdefined\aereproduced
      \includegraphics[width=12mm,height=12mm]{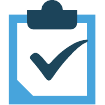}
    \fi
  }
  \fancyfoot[C]{}
}
\begin{document}
\maketitle

\ifdefined\hpcacameraready 
  \thispagestyle{camerareadyfirstpage}
  \pagestyle{empty}
\else
  \thispagestyle{plain}
  \pagestyle{plain}
\fi

\newcommand{\hpcaheight}{0mm}
\ifdefined\eaopen
\renewcommand{\hpcaheight}{12mm}
\fi



\begin{abstract}

This paper studies the impact of DRAM writes on DDR5-based system.  To efficiently perform DRAM writes, modern systems buffer write requests and try to complete multiple write operations whenever the DRAM mode is switched from read to write.  When the DRAM system is performing writes, it is not available to service read requests, thus increasing read latency and reducing performance. We observe that, given the presence of on-die ECC in DDR5 devices, the time to perform a write operation varies significantly: from 1x (for writes to banks of different bankgroups) to 6x (for writes to banks within the same bankgroup) to 24x (for conflicting requests to the same bank).  If we can orchestrate the write stream to favor write requests that incur lower latency, then we can reduce the stall time from DRAM writes and improve performance. However, for current systems, the write stream is dictated by the cache replacement policy, which makes eviction decisions without being aware of the variable latency of DRAM writes. The key insight of our work is to improve performance by modifying the cache replacement policy to increase bank-parallelism of DRAM writes.

Our paper proposes {\em BARD (Bank-Aware Replacement Decisions)}, which modifies the cache replacement policy to favor dirty lines that belong to banks without pending writes. We analyze two variants of BARD: BARD-E (Eviction-based), which changes the eviction policy to evict low-cost dirty lines, and BARD-C (Cleansing-Based), which proactively cleans low-cost dirty lines without modifying the eviction decisions. Both BARD-E and BARD-C increase bank-parallelism of DRAM writes by 30\%. As both versions of BARD have differing impact, extra misses or extra writebacks, there is no single policy that works well across all the workloads. We develop a hybrid policy (BARD-H), which uses a selective combination of both eviction and writeback.  Our evaluations across workloads from SPEC2017, LIGRA, STREAM, and Google server traces show that BARD-H improves performance by 4.3\% on average and up-to 8.5\%. BARD requires only 8 bytes of SRAM per LLC slice.

\end{abstract}

\section{Introduction}

The performance of modern processors is often limited by the latency of accessing the DRAM-based memory system. Even with aggressive out-of-order processors with large instruction windows, the processor is unable to tolerate the long-latency stall produced by the memory access~\cite{mutlu2003runaheadexecution, qureshi2006mlpawarecachereplacement, bera2021pythia, bera2022hermes, bera2024constable}.  The memory accesses caused by load misses (reads) tend to be latency critical, as subsequent instructions often have data dependency on the value of the load instruction. The memory accesses caused by writes are not as latency-critical for processing, as the writeback to DRAM usually happens millions of cycles after the store instruction, often percolating through several levels of write-back caches. While instruction processing does not need to wait for DRAM writes, we show that the latency incurred by performing the DRAM writes can still cause a slowdown for the overall system. Although optimizing memory read latency has received significant attention~\cite{lin2018duplicon, lee2009banklevelparallelismprefetching, lee2013tieredlatencydram, hassan2016chargecache, kaseridis2011mop, navarro2022berti, kim2016spp, bera2021pythia, pakalapati2020ipcp, bera2022hermes, subramanian2016bliss, kim2010parbs, kim2010atlas}, there has not been as much effort in reducing DRAM write-related overheads. The focus of our paper is reducing write-related overheads in DDR5-based memory systems.

\begin{figure*}
    \centering
    \includegraphics[width=0.95\textwidth]{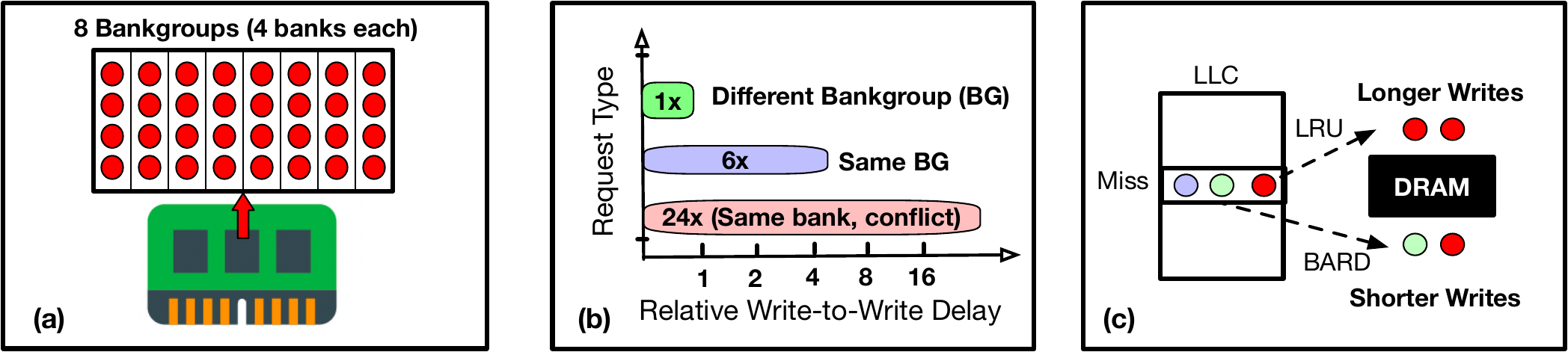}
    \caption{(a)~DRAM chips are organized into backgroups (BG). DDR5 chips have eight BG and a total of 32 banks. (b)~The latency for write depends on the type of request: writes to another bank in a different BG incurs low latency (1x), writes to another bank within the same BG incurs higher latency (6x), and writes to the same bank under a row-buffer conflict incurs much higher latency (24x). (c)~Conventional replacement policy evicts dirty lines without accounting for write latency.  Our proposal, BARD, performs proactive dirty evictions in a bank-aware manner to increase BLP and reduce the latency of writes. }
       \vspace{-0.15 in}

    \label{fig:intro}
\end{figure*}

\vspace{0.05 in}
\noindent{\bf{Evolution of DRAM Architectures:}} The architecture of DRAM chips is decided by the JEDEC specifications.  Over the last decade, the DRAM technology has moved from DDR3 to DDR4 and now to DDR5.  While the access latency of DRAM chips (in terms of nanoseconds) has largely remained unchanged, each successive DRAM generations tries to improve the concurrency and bandwidth of DRAM chips. For example, DDR3 had 8 banks per chip, DDR4 increased it to 16 banks, and DDR5 to 32 banks.  To reduce the circuitry to perform read-write operations in a bank, DDR4 and DDR5, group the banks into {\em bankgroups}. For example, DDR5 chips contain eight bank groups, each containing four banks, as shown in Figure~\ref{fig:intro}(a). The increased complexity of DDR5 specifications has two key implications. First, as the chip contains 32 banks, it offers about 4x more {\em Bank-Level Parallelism} (BLP, the ability to hide bank access-latency by concurrently accessing multiple banks) compared to DDR3 chips. This BLP is enough to fully saturate the DRAM bus, even without exploiting any row-buffer hits.  Second, as the banks in a bank group share some circuitry, consecutive operations to banks within the same bankgroup are often slower than consecutive operations to banks that are in different bankgroups.

\vspace{0.05 in}
\noindent{\bf{Servicing DRAM Writes:}} DRAM modules use a single data bus to perform read and write operations between the processor and the DRAM chips. The data bus is a {\em simplex} connection that can only operate in one direction at a time. Changing the direction of the bus incurs a latency overhead, termed as {\em Bus Turnaround Time}~\cite{chatterjee2012stagedreads, micronddr5sdram, stuecheli2010virtualwritequeue, chang2014refreshparallelwithaccesses, seshadri2014dirtyblockindex, wang2012lastlevelwritepredictor}. As the bus turnaround time can be significant (22ns), memory controllers try to batch multiple write operations and perform them together, thereby incurring only one set of bus turnaround times (read-to-write and then switching back from write-to-read) across all the batched writes. This batching is accomplished by buffering the write requests in a {\em write queue (WRQ)} associated with the DRAM channel~\cite{chatterjee2012stagedreads, rixner2000memory, stuecheli2010virtualwritequeue, wang2012lastlevelwritepredictor, mutlu2007stalltimescheduling, subramanian2016bliss, yuan2009acceleratorscheduling}. The memory controller switches to performing writes when the WRQ occupancy reaches a {\em high watermark} and the WRQ is drained until its occupancy reduces to a {\em low watermark}.

While write operations are typically not considered latency critical, they can still cause performance overhead. When the data bus switches to service writes, DRAM modules cannot service reads. This increases the service time for reads and causes significant slowdown.  Our analysis shows that DRAM modules spend 30\% of the time performing write operations.

\vspace{0.05 in}
\noindent{\bf{Not All Writes are Created Equal:}}  The latency of a write to DRAM depends on the location of the prior write requests due to specific DRAM timing constraints, as in Figure~\ref{fig:intro}(b). For example, if a write request maps to another bank that is located in a separate bankgroup, then it can also be performed with low latency (1x). However, consecutive writes to different banks within the same bankgroup incur a higher latency (6x).

\begin{tcolorbox}
 Note that \textit{row buffer hits also incur this higher 6x write latency, therefore, conventional wisdom of optimizing row-buffer hit-rate for writes~\cite{stuecheli2010virtualwritequeue} is not optimal for DDR5.} 
\end{tcolorbox}

Finally, if a request maps to the same bank but to a different row (row-buffer conflict), then it incurs a significantly higher latency (24x). Generally, the memory controller (MC) tries to issue lower latency writes from the WRQ. However, the WRQ typically contains both low-latency and long-latency write requests. Our study shows that if all writes are performed with low latency (1x), then the time the DRAM system spends on performing writes can be reduced from 33.0\% to 24.1\%.

\vspace{0.05 in}
\noindent{\bf{Insight: High Bank-Parallelism Reduces Write Latency:}} If we can orchestrate the write stream to favor lower latency write requests, then we can reduce the stall time from DRAM writes and improve performance. However, the write stream (buffered in the WRQ) is dictated by the replacement policy of the Last-Level-Cache (LLC), which performs evictions unaware of the variable latency of DRAM writes. If the line maps to a bank with a pending write request, writing the line to DRAM will incur higher latency, as shown in Figure~\ref{fig:intro}(c).

The key insight of our work is to reduce the write latency of DDR5 system by increasing the bank-parallelism of the write stream, as this causes fewer bankgroup and bank conflicts. To this end, we propose {\em BARD (Bank-Aware Replacement Decisions)}, a modification to the LLC replacement policy that improves the bank parallelism of the DDR5 write stream. 

\newpage

{
\noindent{\bf{Tracking Bank-Parallelism with Low Cost:}} To avoid high latency write requests, BARD must track banks with pending writes to avoid sending writes to those banks. A na\"{i}ve strategy is scanning the MC's WRQs during every victim selection. While accurate, this increases on-chip network communication and power consumption and is impractical to implement.

Instead, we develop a simpler design called \textit{BLP-Tracker}, which is \textit{contained entirely in the LLC and does not require any communication with the MC}. The BLP-Tracker uses one bit per bank to indicate whether a bank has a pending write request. The BLP-Tracker sets a bank's bit once the LLC writes back a line mapping to that bank. It also resets itself once all bank bits are set (all ones). The BLP-Tracker only requires 8B of SRAM per channel per LLC slice to implement.


\vspace{0.05 in}
\noindent{\bf{Implementing BARD:}} We explore two variants of BARD. The first, {\em BARD-E (Eviction-Based)}, only operates when the LRU victim is \textit{dirty}. If the dirty victim maps to a bank with a pending write (determined using the BLP-Tracker), BARD scans the set from LRU to MRU for a dirty line which maps to a bank without a pending write. If this line is found, it is evicted instead. The second, {\em BARD-C (Cleansing-Based)}, only operates when the victim is \textit{clean}. BARD searches the set for a dirty line mapping to a bank without a pending write. If one is found, it is cleaned (writeback without eviction).

As BARD-E and BARD-C have differing costs (extra misses vs extra writebacks), neither are universally better across all workloads. We propose {\em BARD-H (Hybrid)} which uses BARD-E when the victim is dirty and uses BARD-C when the victim is clean. BARD-H provides a gmean speedup of 4.3\% and requires 8B of SRAM per channel per LLC slice.

\vspace{0.05 in}
\noindent{\bf{Contributions:}} Our contributions are as follows:
\begin{enumerate}
[leftmargin=0.1cm,itemindent=0.4cm,labelwidth=\itemindent,labelsep=0.05cm, align=left, itemsep=0.1cm, listparindent=0.5cm, topsep=0.1cm]
    \item To the best of our knowledge, this is the first work to argue for increasing bank-parallelism to reduce write overheads. 

    \item We propose {\em BARD} (Bank-Aware Replacement Decisions), a low-cost mechanism that favors the eviction ({\em BARD-E}) or cleansing ({\em BARD-C}) of dirty lines that have low latency.

    \item We propose the {\em BLP-Tracker}, which tracks banks with pending writes and {\em only needs 8B SRAM/channel/LLC slice.}


    \item We propose {\em BARD-H} which combines BARD-E and BARD-C and outperforms both.
    
\end{enumerate}


\ignore{
Flow: What is needed?

- How DRAMs works (bank, row, column)

- Next para DDR5 also has bankgroups, subchannels, slower bus (so it takes 8 cycles). Fundamentally you cannot do a read-write operation at a latency lower than 8 cycles. 

- See the table

- Write handling (buffering WRQ, we have 48 entries). Spends too much time doing writes.  The variable latency (diagram for variable latency)

- Write BLP is low (show diagram).  

- Goal

}

 \begin{figure*}[!htb]
        \centering
        \includegraphics[width=\textwidth]{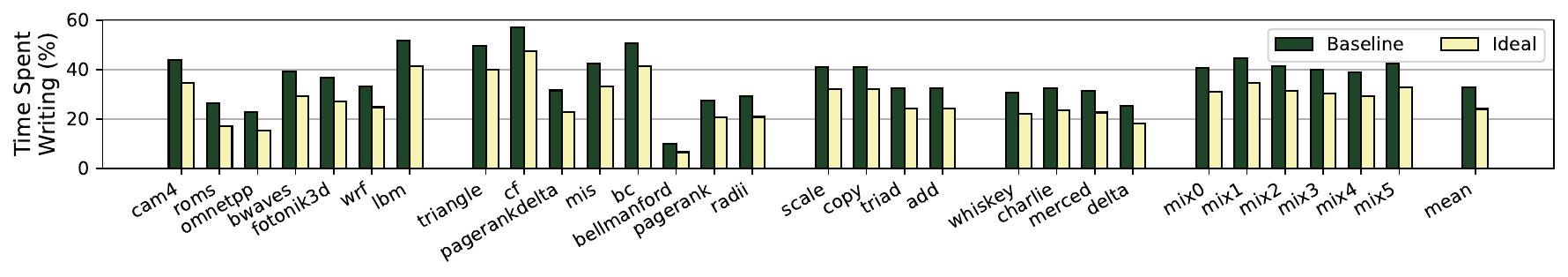}
        \caption{Percentage of total execution spent issuing writes to DRAM for baseline and an idealized system (where a write can be issued every 3.3ns). On average, workloads spend 33.0\% of their time writing to DRAM, and ideally, this can be reduced to 24.1\%.}
        \label{fig:write_mode_time}
    \end{figure*}
\newpage
\section{Background and Motivation}

\subsection{Basic DRAM Organization}

The data in a DRAM module is organized as {\em banks}, consisting of rows and columns. To access the data from a row, the row must first be opened using an {\em activate (ACT)} command.  Then, a read/write operation can be performed on the opened row.  Finally, to obtain data from another row in the same bank, the opened row must be closed using the {\em precharge (RP)} command.  Banks allow concurrency, as the bank access latency can be amortized by performing multiple concurrent requests to different banks. All the banks share a data bus that transfers the data between the processor and DRAM. 



\subsection{DRAM Evolution to DDR5}


While bank access latency has largely remained unchanged, DRAM architectures try to improve memory performance by (a) increasing the transfer rate of the data bus and (b) increasing the concurrency in DRAM modules by increasing the number of banks. For example, DDR3 chips had 8 banks, and the number of banks with DDR5 chips increased to 32.  The salient features of DDR5 are as follows:

\begin{enumerate}
[leftmargin=0.1cm,itemindent=0.4cm,labelwidth=\itemindent,labelsep=0.05cm, align=left, itemsep=0.1cm, listparindent=0.5cm, topsep=0.1cm]
    \item {\bf{Bankgroups:}}  DDR5 chips contain 8 bankgroups, each with 4 banks for a total of 32 banks. Bankgroups were first introduced in DDR4, which had 4 bankgroups.
    
    \item {\bf{Sub-channels:}}  DDR5 divides the chips into two sub-channels, each with a 32-bit data bus. Each sub-channel operates independently. So, a DDR5 rank contains 64 banks.
    
    \item {\bf{Longer Burst-Length:}}  As each sub-channel is 32-bit wide, it requires 16 transfers (8 cycles with DDR) to transfer a 64-byte line.  Thus, each read or write request occupies the bus for eight cycles (up from 4 cycles in DDR3 and DDR4). 
    
    \item {\bf{On-Die ECC:}}  DDR5 modules support on-die ECC to correct single-bit errors transparently within the DRAM chip.
\end{enumerate}

Given the significant increases in the number of banks with DDR5 (32 per sub-channel, 64 per rank), the performance of the system is highly influenced by the ability to exploit the available bank-level parallelism. For example, modern memory mappings, such as AMD-Zen~\cite{jattke2024zenhammer}, split a 4KB page such that it gets distributed across 32 banks (with only two lines of a page co-resident in the same bank).  Such mappings exploit the available {\em bank-level parallelism} for reads, even at the expense of the row-buffer hit-rate.  In fact, with the larger number of banks, it is possible to keep the data bus fully utilized by sending read requests to different banks. 


\subsection{DDR5 Memory Timings}

 The memory controller must obey the DRAM's timing constraints to ensure functional correctness. Table~\ref{tab:ddr5_timing} shows the timing constraints that are most relevant to our study. Note that timing constraints with an \texttt{S} are commands to different bankgroups, whereas those with an \texttt{L} are commands to the same bankgroup. We use x4 devices for our study, as server memories are typically designed with x4 devices (to implement strong reliability, such as Chipkill)~\cite{nair2016xed, jian2013adaptivechipkill}.  
 
\begin{table}[!htb]
    \centering
    \footnotesize
    \begin{center}
    \caption{DRAM Timing (DDR5 4800-B x4 Devices)~\cite{micronddr5sdram}}
    \vspace{-0.2 in}
    \label{tab:ddr5_timing}
    \begin{tabular}{|c||c|c|c|}
        \hline
        \textbf{Name} & \textbf{Description} & \textbf{Time} & \textbf{Cycles}\\
        \hline
        \hline
        $CL$ & Read Latency & 16.6ns & 40\\
        $CWL$ & Write Latency & 15.8ns & 38 \\
        $tRCD$ & Activate-to-RW Latency & 16.6ns & 39 \\
        $tRP$ & Precharge-to-Activate Latency & 16.6ns & 39 \\
        $tRAS$ & Activate-to-Precharge Latency & 32.1ns & 77 \\
        $tWR$ & Write-to-Precharge Latency & 30.4ns & 72 \\
        \hline
        $BL/2$ & Time to send 64B across data bus & 3.3ns & 8 \\
        \hline
        $tCCD\_S\_WR$ & Write-to-Write Delay (Diff.) & 3.3ns & 8 \\
        $tCCD\_L\_WR$ & Write-to-Write Delay (Same) & 20.4ns & 48 \\
        \hline
       
    \end{tabular}
    \end{center}
\end{table}

We note that consecutive writes to the same bankgroup incurs 6x higher latency than to different bankgroups.  With x4 devices, each chip receives only 64-bit data for writing the 64-byte line. However, DDR5 specifies that the chip must maintain on-die ECC at 128-bit granularity (to reduce cost). Therefore, each 64-bit write to the x4 chip internally needs to perform a read-modify-write of whole 128-bit (including the sibling 64-bit word).  These extra operations to maintain on-die ECC tie up the bankgroup and increases latency.

\begin{figure*}[!htb]
    \centering
    \includegraphics[width=\textwidth]{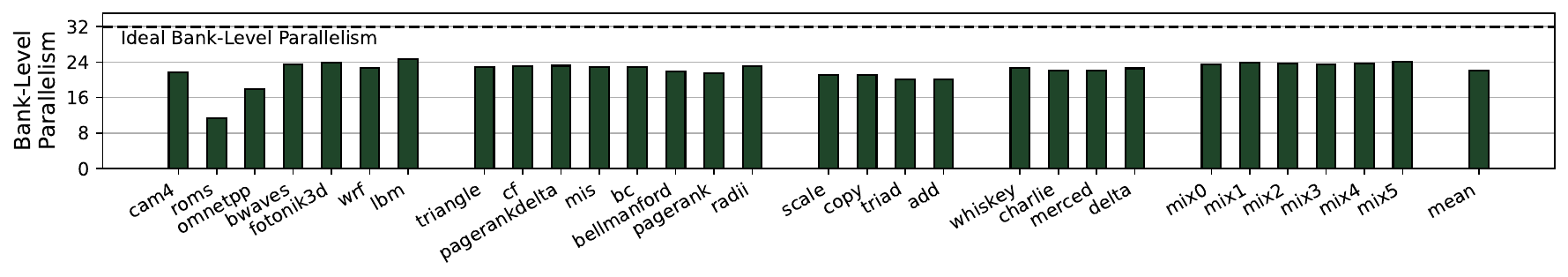}
    \caption{Write bank-level parallelism for all workloads evaluated in this paper. Ideally, workloads should write to all 32 banks within a sub-channel when draining the write buffer. However, on average, workloads write to 22.1 banks.}
    \label{fig:bank_level_parallelism}
\end{figure*}

\subsection{Scheduling DRAM Writes}
\label{sec:write_scheduling}

The data bus used to connect the DRAM modules to the processor is simplex.  Therefore, at any time, it can either service reads or writes, but not both.  To change the mode of operation of the DRAM bus, we need to incur additional latency, called {\em Bus Turnaround Time}. To avoid frequently incurring bus-turnaround penalty, the memory controller buffers the writes in a {\em Write Queue (WRQ)} and tries to process multiple write requests in a batch (thus amortizing the cost of bus turnarounds)~\cite{chatterjee2012stagedreads, rixner2000memory, stuecheli2010virtualwritequeue, wang2012lastlevelwritepredictor, mutlu2007stalltimescheduling, subramanian2016bliss, yuan2009acceleratorscheduling}. Writes occur when the WRQ occupancy reaches a {\em high watermark} and stops when the occupancy drops to a {\em low watermark}.  Although writes are not latency critical, when the DRAM module is performing writes, it cannot service read requests, and the stall time increases read latency and reduces performance. 

\subsection{Variability in Write-to-Write Timings}

When the DRAM bus mode is changed from read to write, the memory controller must be able to perform the write requests stored in the WRQ (up-to the low watermark) within a short amount of time. Unfortunately, the total time required to finish a given number of writes depends on the location of those writes. Given that DRAM write must occupy the data-bus for 8 cycles, we cannot do a write operation at a rate faster than once per 8 cycles.
Figure~\ref{fig:bglatency} shows the latency between consecutive writes if they are distributed across different bankgroups. Each write is completed as fast as possible (8 cycles). However, if a write request maps to the same bankgroup as a prior write, it incurs a latency of 48 cycles (6x longer). Ideally, we want the write traffic to be distributed across a large number of bankgroups (and banks).  
    
  \begin{figure}[!htb]
        \centering
        \vspace{-0.1 in}
        \includegraphics[width=0.75\columnwidth]{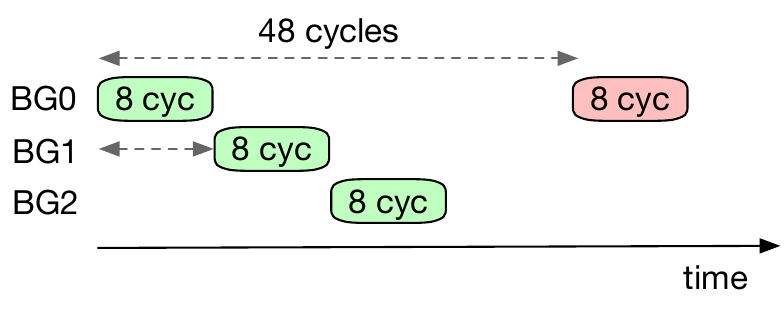}
        \vspace{-0.1 in}
        \caption{Write-to-write delays in DDR5 for another bank located in same bankgroup and different bankgroup. Consecutive writes to same bankgroup incur 6x higher latency.}
        \label{fig:bglatency}
    \end{figure}

If consecutive writes go to the same bank, then latency depends on row-buffer hit or conflict. For row-buffer hits, the latency is the same as writes to same bankgroup (48 cycles). However, in case of row-buffer conflict, we incur the longer latency of write recovery time, precharge, and additional activation, for a total of 188 cycles, which is 23.5x higher latency than writes to different bankgroups. 
   
  \begin{figure}[!htb]
        \centering
        \includegraphics[width=0.75\columnwidth]{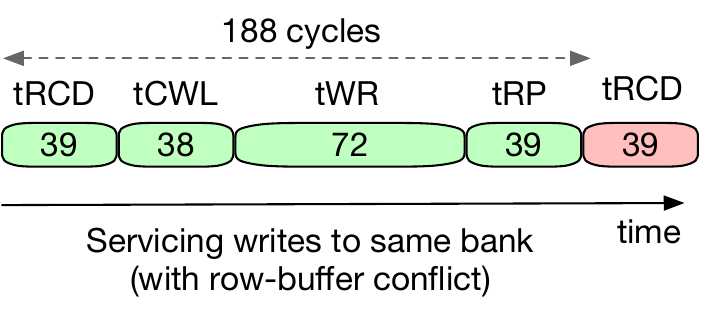}
        \vspace{-0.1 in}
        \caption{Write-to-write delays in DDR5 for serving a row-buffer conflict. The latency is 188 cycles (23.5x higher).}
        \label{fig:conflictlatency}
    \end{figure}

\subsection{Impact of Longer Write-to-Write Timings}

We want all write requests to be performed with minimal write-to-write latency (8 cycles per write), as minimizing write-related overheads reduces read stalls and improves overall system performance. Figure~\ref{fig:write_mode_time} shows the percentage of time the memory system would spend doing DRAM writes if each write occurred with minimal latency (ideal). The stall time due to write operations reduces from 33.0\% to 24.1\%. This shows that there is an opportunity to improve performance if we can steer the write traffic to favor low-latency writes.

One solution is increasing the size of the WRQ. A larger WRQ provides more candidates for scheduling, so the memory controller would be more likely to avoid bank and bankgroup conflicts. Unfortunately, WRQs are large, fully associative structures, so expanding the WRQ would greatly increase area, latency, and power overheads.


Another strategy is to ensure that the write requests arriving at the WRQ map to different bankgroups and banks. 
Unfortunately, in current systems, the write stream arriving at the WRQ stems from the decisions of the LLC's replacement policy. Typically, the LLC optimizes for miss rate, so its decisions are influenced neither by the destination of a writeback nor the pending requests in the WRQs. Figure~\ref{fig:bank_level_parallelism} shows the average number of banks that are accessed during the process of draining a write-queue (for performing 32 writes). Ideally, all requests should map to different banks to exploit bank-parallelism. On average, only 22 out of the 32 banks are used. Thus, conventional systems have low bank-level parallelism for writes, causing high write-to-write delay.


\subsection{Goal}

The goal of our paper is to improve the bank-parallelism for write requests, with the aim of reducing the latency of write operations in DDR5-based memory system. Our key insight in achieving this objective is to modify the cache replacement policy to proactively evict (or clean) the dirty lines that map to banks that do not have a pending write and delay the eviction of dirty lines that would otherwise incur higher latency.


\newpage

\section{Evaluation Methodology}
\label{sec:evalmethodology}

\subsection{Simulation Framework}
    Our evaluations are conducted using ChampSim~\cite{gober2022champsim} extended with a detailed DDR5 memory model. Table~\ref{tab:baseline_config} describes the baseline configuration. We use AMD-Zen~\cite{jattke2024zenhammer} mapping, shown in Figure~\ref{fig:zen_mapping}, with an adaptive open-page policy which closes a row if there are no pending requests to the row. We implement \textit{Permutation-Based Page Interleaving (PBPL)}~\cite{zhang2000pageinterleaving}, a common, lightweight optimization that maps lines within the same LLC set to different DRAM banks by swizzling the bank address bits with a subset of row address bits. PBLP reduces bank conflicts.

    \begin{table}[!htb]
        \vspace{-0.1in}
        \begin{center}
            \footnotesize
            \caption{Baseline System Configuration}
            \vspace{-0.1in}
            \label{tab:baseline_config}
            \begin{tabular}{|c|c|}
                \hline
                OoO Cores & 8 cores at 4GHz, 512 entry ROB \\

                iTLB and dTLB & 16 sets, 4 ways, 8 MSHRs, LRU \\
                L2 TLB & 1024 sets, 12 ways, 16 MSHRs, LRU \\ 
                L1I Cache & 32KB, 8 way, 8 MSHRs, LRU \\
                L1D Cache & 48KB, 12 way, 16 MSHRs, LRU + Berti~\cite{navarro2022berti}\\
                L2 Cache & 512KB, 8 way, 32 MSHRs, LRU + SPP~\cite{kim2016spp} \\
                Last Level Cache & 16MB, 16 way, 128 MSHRs, LRU \\
                
                \hline
                \hline
                DRAM & 32GB, DDR5 4800 MT/s \\
                R/W Queues & 64-Entry RQ, 48-Entry WQ \\
                WQ Watermarks & Low=8, High=40 \\
                Memory Scheduler & FRFCFS~\cite{rixner2000memory} with Read Priority \\
                Channels & 1 (contains 2 sub-channels) \\
                Banks $\times$ Bankgroups & $4 \times 8$ \\
                Row Closure Policy & Adaptive Open-Page \\
                Address Mapping & Zen~\cite{jattke2024zenhammer} + PBPL~\cite{zhang2000pageinterleaving} \\
                \hline
            \end{tabular}
        \end{center}
    \end{table}

\subsection{Evaluated Workloads}
    \label{sec:evaluated_workloads}
    We evaluate using workloads from the SPEC2017, LIGRA~\cite{shun2013ligrabenchmarks}, and STREAM~\cite{mccalpin1995streambenchmarks} benchmark suites and server traces from Google~\cite{dynamoriogoogletraces} whose LLC writebacks per kilo-instruction, or \textit{WPKI}, exceed $2.5$ in the baseline. We also evaluate 6 mixes consisting of 8 different workloads selected from SPEC2017, LIGRA, and Google server traces. All evaluations (8-core ratemode and mixes) first execute 25M warmup instructions followed by 100M simulated instructions. We measure performance using \textit{weighted speedup}.
    
    Table~\ref{tab:mixes} lists the mixes we use in our evaluations. Table~\ref{tab:workloads} shows the LLC misses per kilo-instruction (MPKI), WPKI, write bank-level parallelism (WBLP), and percentage of execution time spent writing (W\%) for each workload.
 
    \begin{table}[!htb]
        \vspace{-0.1in}
        \begin{center}
            \footnotesize
            \caption{Mix (Heterogenous) Workloads}
            \vspace{-0.1in}
            \label{tab:mixes}
            \begin{tabular}{|c|c|}
                \hline
                Mix Number & Constiuent Workloads \\
                \hline
                \hline
                \multirow{2}{*}{mix0} & cam4, omnetpp, lbm, cf \\
                                      & mis, whiskey, merced, delta \\
                \hline
                \multirow{2}{*}{mix1} & roms, bwaves, triangle, pagerankdelta \\
                                      & bc, whiskey, charlie, delta \\
                \hline
                \multirow{2}{*}{mix2} & roms, fotonik3d, wrf, triangle \\
                                      & bc, bellmanford, pagerank, radii \\
                \hline
                \multirow{2}{*}{mix3} & omnetpp, bwaves, cf, pagerankdelta \\
                                      & mis, bellmanford, pagerank, radii \\
                \hline
                \multirow{2}{*}{mix4} & cam4, fotonik3d, wrf, lbm \\
                                      & bc, radii, charlie, merced \\
                \hline
                \multirow{2}{*}{mix5} & roms, bwaves, fotonik3d, wrf \\ 
                                      & lbm, triangle, pagerankdelta, delta \\
                \hline
            \end{tabular}
        \end{center}
    \end{table}
    
    \begin{figure}[!htb]
        \centering
        \includegraphics[width=\columnwidth]{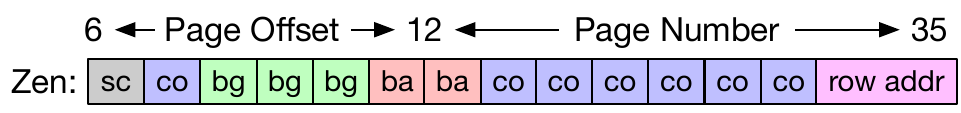}
        \caption{The AMD Zen address mapping~\cite{jattke2024zenhammer}. \texttt{sc}, \texttt{bg}, \texttt{ba}, and \texttt{co} correspond to sub-channel, bankgroup, bank, and column bits.}
        \label{fig:zen_mapping}
    \end{figure}

    \begin{table}[!htb]
        \begin{center}
            \footnotesize
            \caption{Workload Characteristics}
            \vspace{-0.1in}
            \label{tab:workloads}
            \begin{tabular}{|c|c||c|c|c|c|}
                \hline
                Suite & Workload & MPKI & WPKI & WBLP & W\% \\
                \hline
                \hline
                \multirow{7}{*}{SPEC2017}

& cam4            &       9.2 &   4.1 &   21.6 &          43.9 \\
& roms            &      13.2 &   2.7 &   11.4 &          26.3 \\
& omnetpp         &      13.7 &   5.5 &   17.9 &          22.7 \\
& bwaves          &      20.8 &   6.1 &   23.4 &          39.3 \\
& wrf             &      25.4 &   7.3 &   22.7 &          33.1 \\
& fotonik3d       &      30.6 &   9.7 &   23.9 &          36.9 \\
& lbm             &      48.5 &  25.5 &   24.6 &          51.8 \\

                \hline
                \hline
                \multirow{8}{*}{LIGRA~\cite{shun2013ligrabenchmarks}}

& triangle        &      15.9 &   8.1 &   22.8 &          49.6 \\
& pagerankdelta   &      25.3 &   8.1 &   23.2 &          31.6 \\
& mis             &      26.1 &  10.4 &   22.8 &          42.3 \\
& bellmanford     &      45.2 &   3.3 &   21.9 &          10.1 \\
& cf              &      48.3 &  16.2 &   23.1 &          57.3 \\
& bc              &      57.2 &  20.7 &   22.9 &          50.6 \\
& radii           &      60.7 &  16.0 &   23.1 &          29.3 \\
& pagerank        &      70.0 &  10.9 &   21.4 &          27.4 \\

                \hline
                \hline
                \multirow{4}{*}{STREAM~\cite{mccalpin1995streambenchmarks}}

& triad           &      110.8 &         18.5 &   20.1 &          32.3 \\
& scale           &      123.8 &         21.0 &   21.2 &          40.9 \\
& copy            &      128.2 &         26.4 &   21.1 &          41.0 \\
& add             &      129.3 &         21.7 &   20.1 &          32.3 \\

                \hline
                \hline
                \multirow{4}{*}{Google Srv.~\cite{dynamoriogoogletraces}}

& charlie         &      16.1 &   5.3 &   22.0 &          32.4 \\
& whiskey         &      19.2 &   5.1 &   22.7 &          30.8 \\
& merced          &      20.0 &   5.7 &   22.2 &          31.3 \\
& delta           &      27.3 &   5.1 &   22.6 &          25.4 \\

                \hline
                \hline
                \multirow{6}{*}{Mixes}

& mix0            &      41.4 &  17.9 &   23.4 &          40.8 \\
& mix1            &      48.3 &  29.1 &   24.0 &          44.7 \\
& mix2            &      49.6 &  21.8 &   23.7 &          41.3 \\
& mix3            &      64.9 &  29.6 &   23.4 &          40.0 \\
& mix4            &      87.6 &  32.5 &   23.7 &          38.8 \\
& mix5            &      91.1 &  44.7 &   24.0 &          42.6 \\

                \hline
            \end{tabular}
            \vspace{-0.2in}
        \end{center}
    \end{table}

\ignore{
\begin{figure*}[!t]
    \centering
    \includegraphics[width=\textwidth]{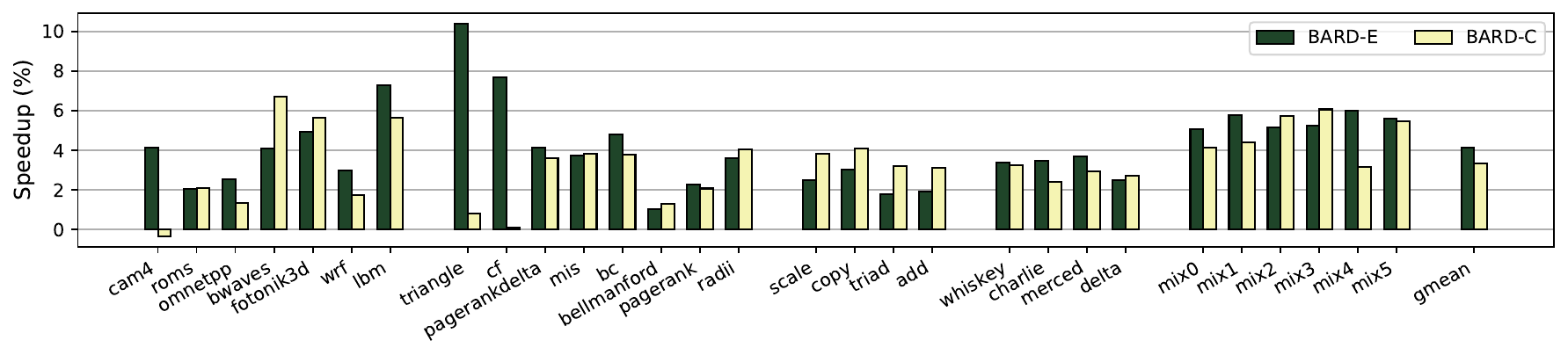}
    \vspace{-0.3 in}
    \caption{Speedups of BARD-E and BARD-C over the baseline. BARD-E improves performance by 3.5\% on average, and BARD-C improves performance by 4.0\% on average. Neither policy outperforms the other on \textit{all} workloads.}
    \vspace{-0.1 in}
    \label{fig:bard_ec_comparison}
\end{figure*}}
\section{Bank-Aware Replacement Decisions (BARD)}

As DRAM write latency depends on how pending write requests map to the DRAM banks, we want to reduce the performance impact of DRAM writes by steering the system toward performing write operations with lower latency. The key insight in our work is to increase the {\em Bank-Level Parallelism (BLP)} of write operations, where BLP denotes the number of unique banks that receive writes in one write episode. For example, if there are 32 pending writes, and they all map to different banks, then all 32 writes will incur minimal write-to-write latency, thus reducing write-related stalls.

To this end, we propose {\em Bank-Aware Replacement Decisions (BARD)}, a modification to the replacement policy (and writeback policy) of the last-level cache, with the goal of increasing BLP of write requests. Ordinarily, the cache replacement policy aims to reduce misses for the last-level cache. It does not focus on modulating the write traffic. BARD modifies the replacement policy to specifically orchestrate the writeback traffic to the WRQ, such that the WRQ receives write requests from as many unique banks as possible (thus increasing the BLP for writes). We develop two variants of BARD: one that focuses on modifying the evictions (BARD-E) and one that proactively writebacks dirty lines (BARD-C).

\subsection{BLP-Tracker: a Low-Cost Write BLP Tracker}
\label{sec:blp_tracker}
BARD requires a mechanism to determine whether a bank has a pending write request. A na\"{i}ve strategy would be probing the memory controller's WRQs to check if a given bank has any pending writes every victim selection. While accurate, this is infeasible for two reasons. First, this requires leveraging the WRQ's forwarding logic and would greatly increase power consumption. Second, frequent communication between the various LLC banks/slices and the WRQ (at the memory controller) may introduce a significant amount of traffic on the on-chip network.

Given these downsides, we develop a \textbf{design that is contained within the LLC and does not require any communication with the WRQ}. To this end, we propose the \textit{BLP-Tracker}. Each DRAM bank has a corresponding bit in the BLP-Tracker which indicates whether the bank has recently received a writeback. If the bit is set, BARD assumes that a bank has a pending write request.
Figure~\ref{fig:blp_vector_impl}(a) describes how BARD accesses a dirty line's corresponding BLP-Tracker bit. First, BARD uses the DRAM address mapping function to compute the line's bank-id. This bank-id is used to index into the BLP-Tracker and access the bank's bit, allowing BARD to read the bit to determine if a line improves BLP, or set the bit after the LLC issues a writeback. 
Figure~\ref{fig:blp_vector_impl}(b) describes the BLP-Tracker's reset mechanism. \textbf{The BLP-Tracker resets itself:} if the bank bits for a sub-channel are all ones, the BLP-Tracker resets all bits to $0$.

\begin{figure}[!htb]
    \centering
    \includegraphics[width=0.9\columnwidth]{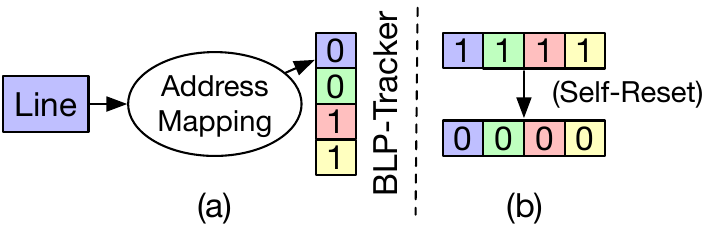}
    \caption{(a)~BARD uses the DRAM address mapping function to index into the BLP-Tracker. (b)~BLP-Tracker is self resetting: Once all bits of sub-channel are 1, the BLP-Tracker self-resets.}
    \label{fig:blp_vector_impl}
\end{figure}

For simplicity, we elect to have each LLC slice maintain its own BLP-Tracker, and that all BLP-Trackers are synchronized. As the BLP-Tracker is only updated when a writeback occurs, the LLC must broadcast a line's bank-id across the on-chip network following a writeback. As a bank-id is a few bits (6b in our setup), broadcasting the bank-id will not noticeably increase network traffic.

\subsection{BARD-E: Eviction-Based BARD}

Figure~\ref{fig:barde} shows an overview of BARD-E.  Without loss of generality, we assume a cache design that uses LRU replacement policy\footnote{See Section~\ref{sec:other_repl_policies} which discusses BARD with other replacement policies.}. Normally, on a cache miss, the replacement policy would select the LRU line as victim. If the victim is dirty, the cache will insert the line into WRQ. Instead, if the LRU victim is dirty, BARD-E first checks if its eviction improves BLP by accessing the victim's BLP-Tracker bit. If the victim's BLP-Tracker bit is $0$, then the line is evicted as the victim's DRAM bank has yet to receive a writeback. Otherwise, BARD-E scans the remaining ways in the set, from LRU to MRU, for a dirty line that improves BLP (checked using the BLP-Tracker). If no such line is found, BARD-E defaults to the LRU victim.
For example, in Figure~\ref{fig:barde}, BARD-E skips the ``red" cache-line, even though it is LRU, as the BLP-Tracker entry corresponding to it is 1, indicating pending write to the red bank. Similarly, the ``blue" cache-line is also skipped (cache-lines for both red and blue banks are present in the WRQ). Finally, the BLP-Tracker for the ``green" cache-line is zero, so it is evicted and placed in the WRQ.

\begin{figure}[!htb]
    \centering
    \includegraphics[width=0.8\columnwidth]{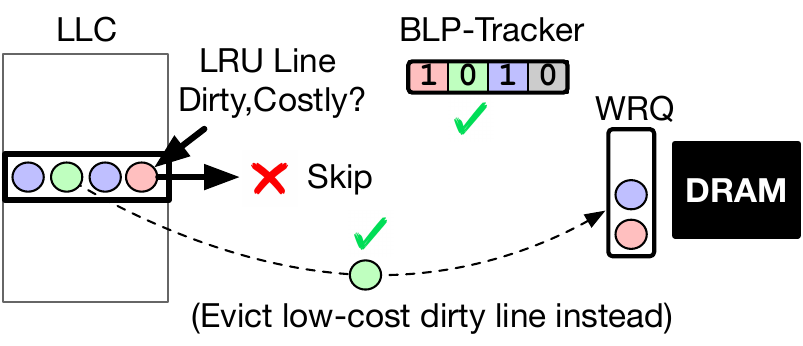}
    \caption{Overview of BARD-E. If LRU line is dirty and to a bank with pending write, the BARD-E selects another dirty line from set that maps to bank with no pending write. }
    \label{fig:barde}
\end{figure}

Note that as BARD-E can make suboptimal replacement decisions (due to the focus on reducing the write cost), it can sometimes cause a slight degradation in performance due to the increase in misses.  Our next policy does not affect the cache miss-rate. 

\subsection{BARD-C: Cleansing-Based BARD}

BARD-C focuses on cleansing, which happens by proactively writing back a dirty line (without evicting it from the cache), thus converting a dirty line into a clean line.  Figure~\ref{fig:bardc} shows an overview of BARD-C. On a miss, if the replacement policy (LRU) selects a clean victim, then BARD-C traverses from LRU to MRU to find a dirty line that maps to a bank with no pending write. If it finds such a line, it copies the content of the line and inserts a request into the WRQ.  It resets the dirty bit associated with the line.

\begin{figure}[!htb]
    \centering
    \includegraphics[width=0.8\columnwidth]{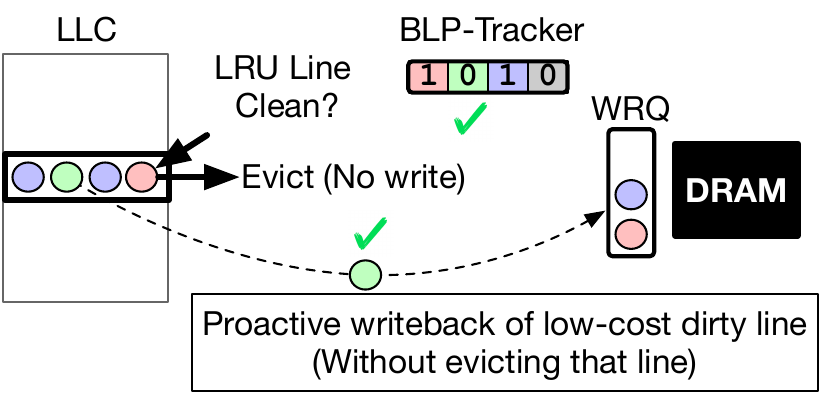}
    \vspace{-0.1 in}
    \caption{Overview of BARD-C. If LRU line is clean, BARD-C tries to do a proactive writeback of low-cost dirty line. }
    \label{fig:bardc}
\end{figure}

\begin{figure*}[!t]
    \centering
    \includegraphics[width=\textwidth]{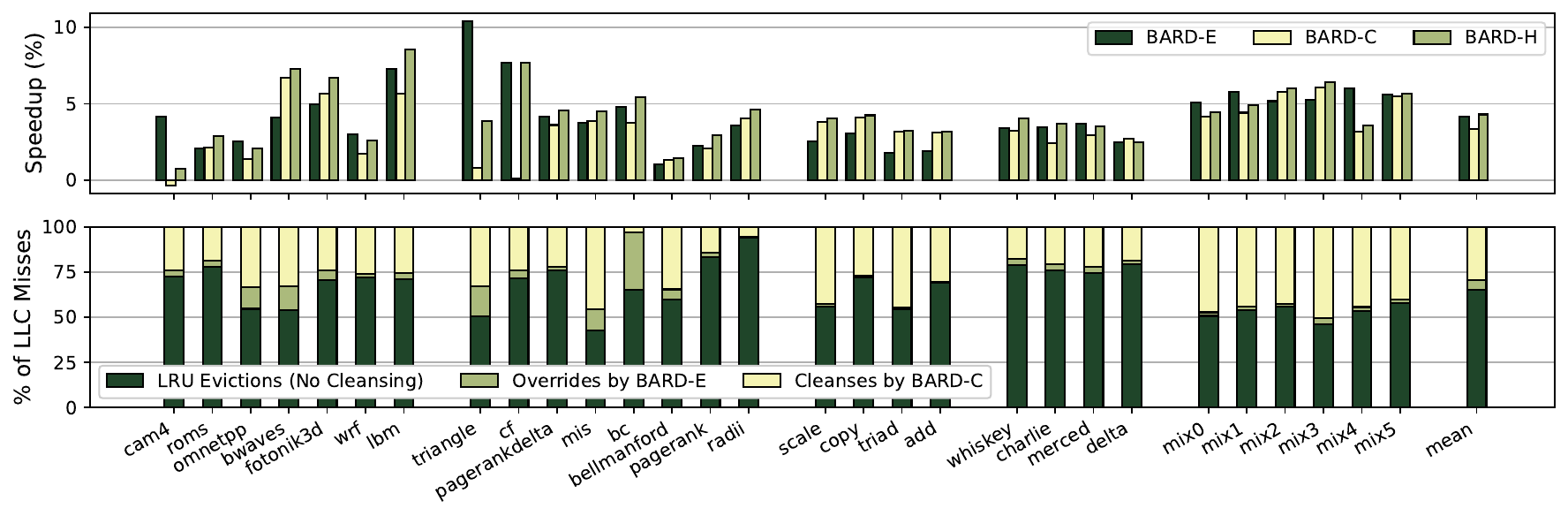}
    \caption{(Top)~Speedups for BARD-E, BARD-C, and BARD-H over the baseline. BARD-E, BARD-C, and BARD-H achieve geomean speedups of 4.1\%, 3.3\%, and 4.3\%, respectively.\\
    (Bottom)~A breakdown of BARD-H's evictions that were (1)~LRU evictions (unaffected by BARD-E or BARD-C), (2)~overrided by BARD-E, and (3)~involved cleaning a dirty line with BARD-C. On average, 64.7\% of evictions are LRU evictions, 4.8\% of evictions were overriden by BARD-E, and 30.5\% of evictions involve cleaning a dirty line with BARD-C.}
    \label{fig:bard_ech_comparison}
\end{figure*}

Like BARD-E, BARD-C relies on the BLP-Tracker to quickly check if the dirty line maps to a bank with no pending writes.  The checking, operation, and resetting of the BLP-Tracker are identical for BARD-E and BARD-C. For the example shown in Figure~\ref{fig:bardc}, on the LLC-miss, the replacement policy identifies the ``red" cache-line as the victim.  It happens to be a clean line, so it is evicted silently (without the need to do any DRAM write).   BARD-C traverses the set from LRU to MRU, and encounters the blue cache-line (dirty) but skips cleansing it, as there is a pending write to the blue bank in the WRQ (as shown by the BLP-Tracker). When BARD-C encounters the ``green" cache-line (dirty), it performs a proactive writeback by placing a write request for that line in the WRQ (and setting the corresponding BLP-Tracker bit).  BARD-C also resets the dirty bit associated with the green cache-line (the line remains in the cache).

While BARD-C has a nice property of not affecting the cache miss-rate, it suffers from two drawbacks: (1)~it is effective only if the victim line is clean, thus it does nothing when the victim is dirty and (2)~it can cause increase in writebacks if the line selected for cleansing receives another write while it is resident in the LLC.

\subsection{Performance of BARD-E and BARD-C}
\label{sec:bard_e_c_comparison}

Figure~\ref{fig:bard_ech_comparison} (top) shows the performance improvement from BARD-E and BARD-C.
On average, BARD-E provides 4.1\% speedup and BARD-C provides 3.3\% speedup.  Overall, neither policy outperforms the other on all workloads. BARD-E and BARD-C have different types of overheads (extra misses versus extra writebacks), so depending on the workload, either policy can be more suited.
 
\section{BARD-H: Hybrid of Eviction and Cleansing}

Both BARD-E and BARD-C are complementary policies, and their costs are different. BARD-C operates only when the LRU victim is clean, and BARD-E operates only when the LRU victim is dirty. So, for a subset of evictions, neither policy would have any scope for improvement.  Ideally, we want to optimize for all evictions regardless of the dirty-bit status of the LRU line.  Our proposal, BARD-H, is a hybrid policy that combines opportunities for both evictions and cleansing.


\subsection{BARD-H: Design}

On a cache miss, the replacement policy selects the victim candidate (LRU line). If that line is dirty, BARD-H uses BARD-E to possibly override the eviction candidate. Otherwise, BARD-H uses BARD-C for proactive cleansing of a dirty line that maps to a bank that has no pending writes. In both cases, we use the BLP-Tracker to check if the bank has any pending writes. 


\revignore{

\begin{figure}[!htb]
        \centering
    \vspace{-0.05 in}
        \includegraphics[width=0.9\columnwidth]{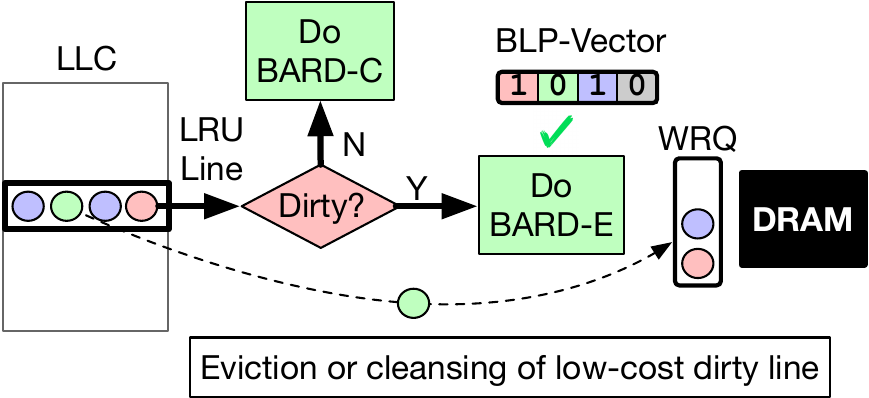}
        \vspace{-0.05 in}
        \caption{Overview of BARD-H (hybrid scheme). If LRU line is dirty, it does BARD-E.  Else, it does BARD-C. }
        \vspace{-0.15 in}
        \label{fig:bardh}
\end{figure}

For the example in Figure~\ref{fig:bardh}, on a cache miss, the replacement policy selects the ``red" cache-line for eviction. BARD-H checks if the evicted cache-line is dirty: (1) If dirty, then BARD-H tries to find another line in the set that is dirty and maps to a bank that has no pending writeback (so, it evicts the ``green" cache-line).   (2) If clean, then BARD-H evicts the ``red" cache-line silently, as there is no need to write it back.  However, it also looks for another dirty line, which maps to a bank with no pending write and cleanses it (so, it cleans the green line).
}

\subsection{Impact on Performance}
    Figure~\ref{fig:bard_ech_comparison} (top) shows the performance improvement of BARD-E, BARD-C, and BARD-H. On average, BARD-E, BARD-C, and BARD-H achieve speedups of 4.1\%, 3.3\%, and 4.4\%. In all cases, BARD-H performs close to the better-performing policy between BARD-E and BARD-C. In some cases (e.g., \texttt{bwaves} or \texttt{lbm}), BARD-H outperforms both BARD-E and BARD-C. 
    

\subsection{Dissecting BARD-H Decisions}
    Figure~\ref{fig:bard_ech_comparison} (bottom) shows BARD-H decisions for each cache miss, dividing them into three categories: (1)~LRU evictions (without cleansing), (2)~evictions overridden by BARD-E, and (3)~evictions involving cleansing with BARD-C. As shown, 64.7\% of evictions are LRU evictions, 4.8\% of evictions are overridden by BARD-E, and 30.5\% of evictions involve cleansing with BARD-C. Note that the disparity between cleanses and overrides is expected as BARD-C removes dirty lines from the set. So, the number of clean LRU evictions increases, which reduces the number of overrides done by BARD-E. Thus, BARD-H primarily uses BARD-C to clean lines from the set. If BARD-C fails to clean any lines before they reach the LRU position, then BARD-E acts as a fail-safe that prevents these dirty lines from degrading BLP.

\begin{figure*}[!t]
    \centering
    \includegraphics[width=\textwidth]{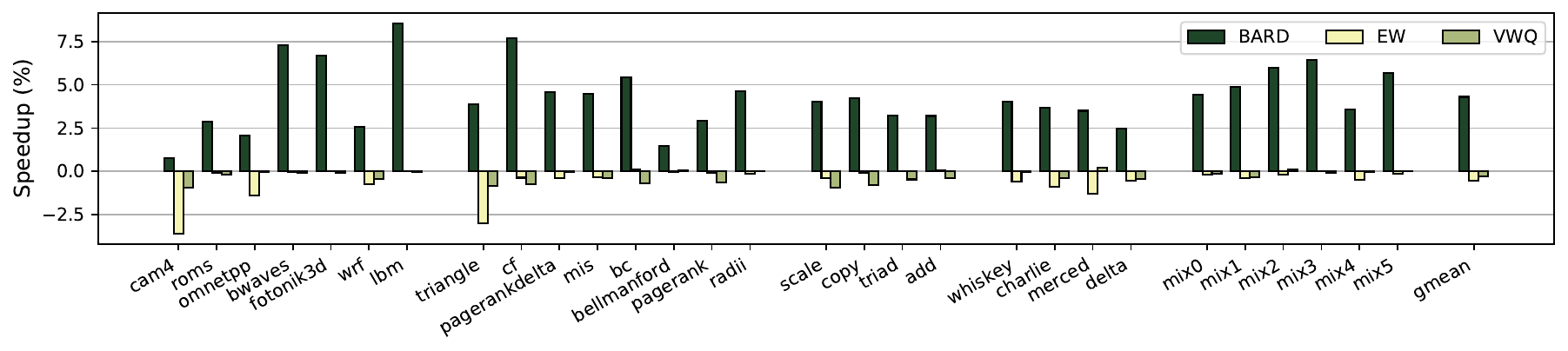}
    \caption{Speedup of BARD, Virtual Write Queue (VWQ), and Eager Writeback (EW). BARD (increased BLP) improves average performance by 4.3\%. EW, with bank-unaware writes, causes 0.5\% slowdown. VWQ, which increases row-buffer hits, causes a 0.3\% slowdown.}
    \label{fig:bard_vwq_ew}
\end{figure*}

\newpage
\section{Comparison with Proactive Writeback}
\label{sec:vwq}

BARD performs proactive writeback of non-LRU dirty lines to improve write performance. The concept of proactive writeback has been applied in prior works to improve the write performance of DRAM~\cite{lee2000eagerwriteback, stuecheli2010virtualwritequeue, seshadri2014dirtyblockindex}. However, prior proposals were developed for older DRAM generations (DDR-DDR3) when the concept of bankgroups (and associated write delays) did not exist. We show that these proposals are ineffective when applied for DDR5.

\subsection{Eager Writeback}

{\em{Eager Writeback (EW)}}~\cite{lee2000eagerwriteback} introduced the concept of eagerly (proactively) performing writebacks of dirty lines (while retaining such lines in the cache). The key insight of EW was to exploit periods of low memory activity to proactively do writebacks, so that writebacks do not impede read requests during periods of high utilization.  Figure~\ref{fig:ew} shows the overview of Eager Writeback.

\begin{figure}[!htb]
        \centering
        \includegraphics[width=0.8\columnwidth]{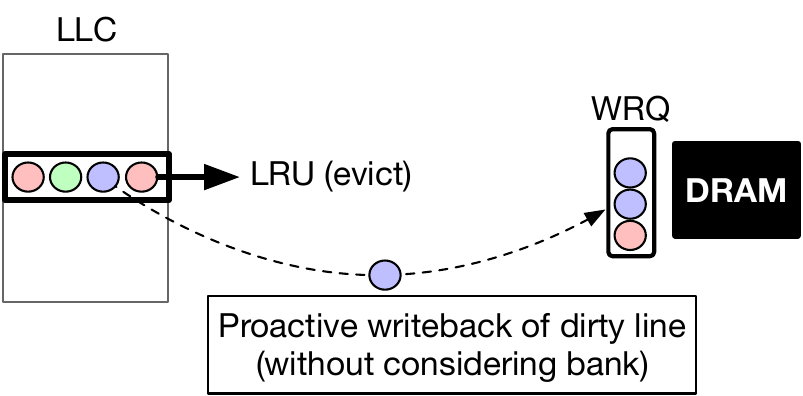}
        \caption{Overview of Eager Writeback. It does proactive writebacks (bank-unaware). It can worsen bank imbalance.}
        \label{fig:ew}
\end{figure}

Eager Writeback suffers from two shortcomings. First, it was evaluated for a single-core system, where phases of a single application can cause the memory system to have a period of low activity.  However, for modern systems with many cores, the likelihood of a channel not receiving any requests for long periods of time is quite low.  Second, Eager Writeback proactively cleans dirty lines in a bank-unaware manner.  For the example in Figure~\ref{fig:ew}, even though the WRQ already has two write requests for ``blue" bank, EW can still do proactive writeback for another blue line, thus worsening the bank imbalance of WRQ entries.

\subsection{Virtual Write Queue}

{\em{Virtual Write Queue (VWQ)}}~\cite{stuecheli2010virtualwritequeue} focuses on increasing the row-buffer hit rate for write requests by proactively doing writeback of dirty lines that map to rows with a pending write request.  Figure~\ref{fig:vwq} shows an overview of VWQ.  On a cache miss, it evicts the LRU line. If this line is dirty, it probes neighboring sets to find a dirty line that maps to the same row as the evicted line. If so, it proactively writes back such lines to get row-buffer hits for write.

\begin{figure}[!htb]
        \centering
        \includegraphics[width=1\columnwidth]{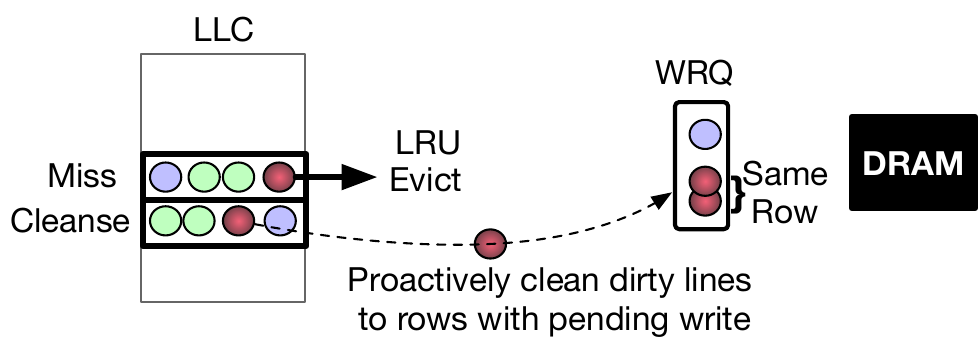}
        \caption{Overview of Virtual Write Queue. It does proactive writebacks of dirty lines to rows with pending writes.}
        \label{fig:vwq}
\end{figure}

VWQ was evaluated for DDR3 (8 banks, no bankgroups). The effectiveness of VWQ is limited for DDR5 for two reasons. First, servicing writes with row-buffer hits still incurs a long latency (6x longer than writes to different bankgroups). Second, given the large number of banks in DDR5 (64 per channel), modern memory mappings (such as AMD-Zen~\cite{jattke2024zenhammer, kaseridis2011mop}) try to exploit BLP for reads by spreading the page across many banks and keeping only 2-4 lines from a page into the same row, limiting row-buffer hits.

\begin{figure*}[!t]
    \centering
    \includegraphics[width=\textwidth]{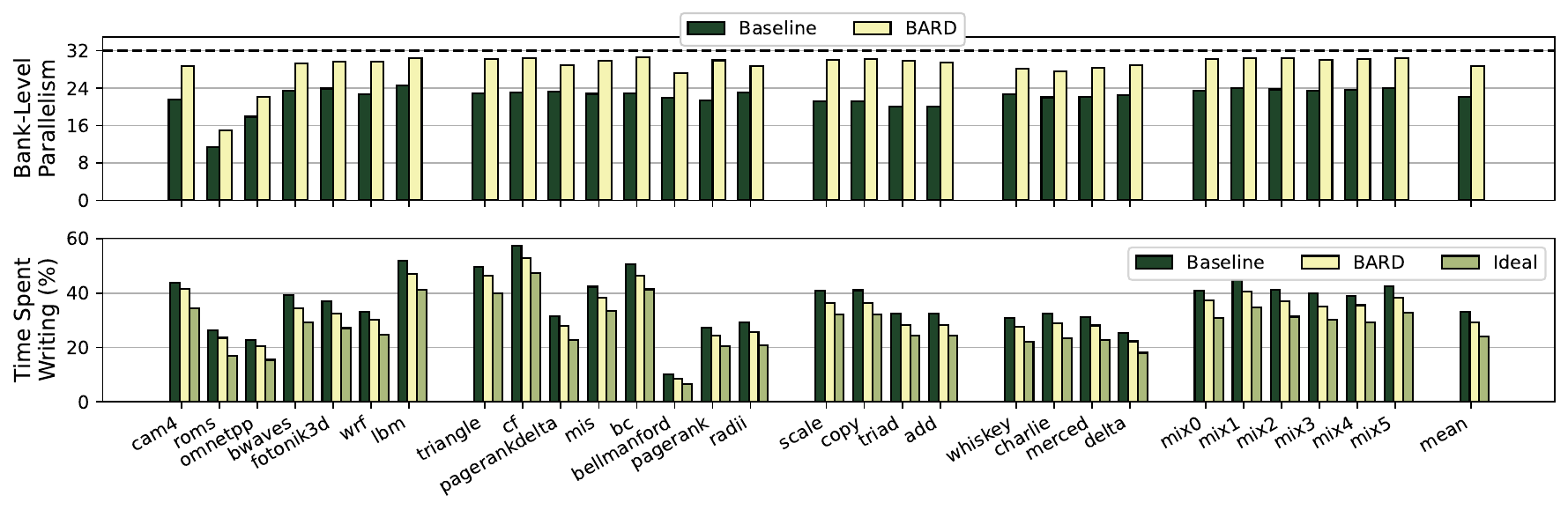}
    \vspace{-0.25in}
    \caption{(Top): Write bank-level parallelism, or the number of banks that perform a write per each episode of  write drain, for the baseline and BARD. The average BLP for writes for the baseline is 22.1, whereas BARD increases it to 28.8 ($1.3\times$ higher).
    \\ (Bottom): Percentage of total execution time spent writing to DRAM. On average, baseline spends 33.0\% of time in writing, and BARD reduces it to 29.3\%. An ideal policy (with all minimum latency writes) would reduce it to 24.1\%.}
    \label{fig:write_blp_comparison}
    \vspace{-0.2in}
\end{figure*}

\vspace{0.1 in}
\subsection{Speedup from Proactive Writeback}
\label{sec:speedup_vs_ew_vwq}

Figure~\ref{fig:bard_vwq_ew} shows the performance improvement with BARD-H, EW, and VWQ. For EW, we writeback the LRU line if it is dirty (without considering the bank) following an eviction or a hit, as these modify the LRU state of the set. For VWQ, we permit it to search the entire LLC for row buffer hits, given that its original design does not work with page-interleaving optimizations used in real systems~\cite{zhang2000pageinterleaving}. On average, BARD-H improves performance by 4.3\%. In contrast, EW and VWQ cause slowdowns of 0.5\% and 0.3\%, respectively.

EW causes slowdowns because it may proactively writeback to banks with pending write requests, thus increasing the number of long latency write requests the memory controller must serve. 
VWQ causes slowdowns because row buffer hits are suboptimal in DDR5 (6x longer latency). Furthermore, as the WRQ has limited capacity, row buffer hit rate and bank-level parallelism are often at odds. Increasing row buffer hit rate requires having multiple entries in the WRQ map to the same row in the same bank, which reduces bank-level parallelism as fewer writes map to different banks. We observe this phenomenon in all workloads where VWQ causes slowdowns. For example, in \texttt{cf}, VWQ increases the row buffer hit rate from 20\% to 28\%. However, the additional row buffer hits reduce bank-level parallelism from 23.3 to 21.9. Thus, VWQ causes a 1\% slowdown. We emphasize that we expect similar results for all prior works that optimize for row buffer hits due to the increased latency of row buffer hits in DDR5~\cite{lee2010dramawarewriteback, seshadri2014dirtyblockindex, wang2012lastlevelwritepredictor}.




\section{Results and Analysis}

For the remainder of this paper, we use BARD to refer to BARD-H. In this section, we examine the impact of BARD on bank-level parallelism, write latency, etc.

\subsection{Impact on Bank-Level Parallelism}

The central premise of the BARD design is to reduce the performance impact of writes by increasing the {\em Bank-Level-Parallelism (BLP)} for write requests. We define BLP for write requests as the number of banks that receive a write request during a given episode of draining of the WRQ. Given the watermarks (high=40, low=8), each write episode tries to service 32 requests; therefore, the maximum BLP for the system would be 32 (the sub-channel contains 32 banks), and we would be able to do each write with the lowest latency (1x).  We also note that trying to maximize the BLP also implicitly increases the bank-group-level parallelism of writes.

Figure~\ref{fig:write_blp_comparison} (top) shows the average BLP for write requests for the baseline and BARD. On average, the baseline has a BLP of 22.1. BARD increases the BLP for writes to 28.8 (1.3x). The increase in bank parallelism for writes is the main reason for BARD's performance improvement. For particularly write-intensive workloads, such as \texttt{lbm} and STREAM workloads, BARD improves BLP up-to 30.6 banks per write drain. While BARD does not achieve perfect bank-level parallelism, note that even minor increases in bank-level parallelism help mitigate write overhead (such as hiding the latency of a write request that experiences a bank conflict).

\subsection{Impact on Time Spent in DRAM Writes}

The increased BLP of BARD reduces the time it takes to perform writes when draining the WRQ, and this equivalently reduces the fraction of system time that is spent doing DRAM writes. Figure~\ref{fig:write_blp_comparison} (bottom) shows the percentage of execution time spent writing to DRAM for the baseline, BARD, and an idealized system where each DRAM write has a latency of 3.3ns. The baseline spends 33.0\% of its execution time writing to DRAM, and the idealized system spends 24.1\% of its execution time writing to DRAM. In contrast, BARD spends 29.3\% of its execution time writing to DRAM, bridging almost half the gap between the baseline and the ideal system.

\subsection{Impact on Write-to-Write Delay}

The implicit assumption of BARD is that increasing BLP would reduce the write-to-write latency for write requests. Table~\ref{tab:dram_write_latency} shows the mean write-to-write latency for the baseline, BARD, and the ideal policy (always doing write at 3.3ns). BARD reduces the average write-to-write delay of the baseline from $5.0$ns to $4.2$ns ($1.2\times$ lower). Thus, BARD is effective in reducing the write-to-write delay. 

\begin{table}[!htb]
    \centering
    \begin{center}
    \caption{Write-to-Write Delay}
    \vspace{-0.1 in}
    \label{tab:dram_write_latency}
    \begin{tabular}{|c||c|c|}
        \hline
        \textbf{Design} & \textbf{Average Latency} & \textbf{Max Latency} \\
        \hline
        \hline
        Baseline & 5.0ns & 5.7ns\\
        \hline
        BARD & 4.2ns & 5.0ns\\
        \hline
        \hline
        Ideal & 3.3ns & 3.3ns\\
        \hline
    \end{tabular}
    \vspace{-0.15 in}
    \end{center}
\end{table}

\begin{figure*}[!t]
    \centering
    \includegraphics[width=\textwidth]{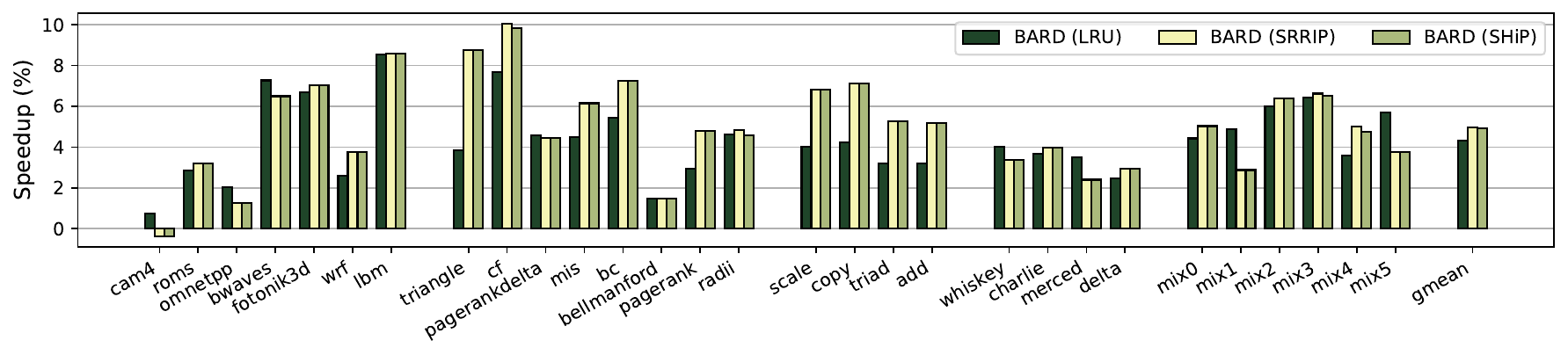}
    \vspace{-0.1in}
    \caption{Performance impact of BARD when the cache uses LRU, SRRIP, or SHiP replacement. On average, BARD provides improvements of 4.3\%, 5.0\%, and 4.9\% for LRU, SRRIP, and SHiP, respectively. So, BARD is effective in both cases.}
    \vspace{-0.1in}
    \label{fig:srrip_results}
\end{figure*}

\subsection{Impact of x8 DDR5 Modules}

We use a baseline with x4 DIMMs, as x4 DIMMs are used in servers (for implementing Chipkill). DDR5 requires that each chip maintain on-die ECC at a 128-bit granularity. In x4 DIMMs (8 chips per sub-channel), when the memory controller issues a write, each chip receives only 64 bits of data that must be written to part of the larger 128-bit codeword. So, each chip must perform a read-modify-write, thus causing $tCCD\_L\_WR = 20\mathrm{ns}$. However, in x8 DIMMs (4 chips per sub-channel), each chip receives 128 bits of data. Instead of performing a read-modify-write, the chip can overwrite all 128 bits and update ECC accordingly. Consequently, $tCCD\_L\_WR$ reduces to 10ns (still 3x longer than minimum write-to-write delay).

Table~\ref{tab:x8cmp} presents the relative performance of baseline, BARD, and Ideal (doing all writes at lowest latency) for both x4 and x8 DIMMs, relative to an x4 baseline. With x4 DIMMs, the baseline and BARD have speedups of 0.0\% and 4.3\%. With x8 DIMMs, these improvements are 2.1\% and 7.1\%.

\begin{table}[!htb]
    \centering
    \begin{center}
        \vspace{-0.1 in}
    \caption{Relative performance with x4 and x8 devices}
    \label{tab:x8cmp}
    \begin{tabular}{|c||c|c|}
       \hline
       \textbf{System} & \textbf{x4 Device} & \textbf{x8 Device} \\
       \hline
       \hline
       Baseline & 0.0\%  (norm) & 2.1\% \\
       \hline
       BARD & 4.3\% & 7.1\% \\
       \hline
       \hline
       Ideal & 14.5\% & 14.5\% \\
       \hline
    \end{tabular}
    \end{center}
        \vspace{-0.15 in}
\end{table}

\subsection{Impact of Replacement Policy}
\label{sec:other_repl_policies}

Our study uses the LRU replacement policy. LRU has a nice property where all lines are precisely ordered (from MRU to LRU) based on access recency. BARD checks resident lines from LRU to MRU to obtain low-cost dirty lines. However, the insights of BARD are applicable to other policies as well. In this section, we evaluate BARD with RRIP (Re-Reference Interval Prediction)~\cite{jaleel2010srrip}. 

\begin{figure}[!htb]
        \centering
        \vspace{-0.15in}
        \includegraphics[width=0.75\columnwidth]{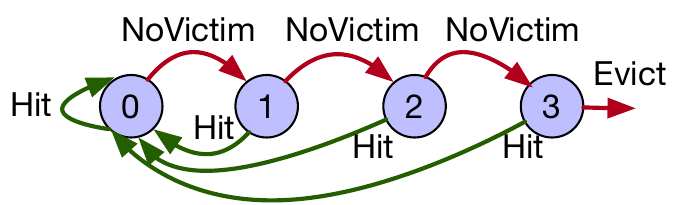}
        \vspace{-0.05 in}
        \caption{Overview of RRIP. On hit, RRPV=0. On miss, line with RRPV=3 is victim (all RRPV incremented if no victim).}
        \label{fig:rrip}
\end{figure}

Figure~\ref{fig:rrip} shows the overview of RRIP. RRIP maintains a 2-bit saturating counter per LLC way: these counters are known as \textit{re-reference predictor values (RRPVs)}. For victim selection, line with RRPV equal to $3$ (the maximum) is evicted, with ties broken arbitrarily. If none of the lines within a set have an RRPV of $3$, then all RRPVs are incremented until one reaches 3. Thus, instead of a precise order, RRIP groups lines into four "buckets" based on RRPV. 

Implementing BARD with RRIP requires minor modifications. On a miss, the cache uses RRIP to select the victim. If the victim is clean, then BARD searches for a dirty line whose BLP-Tracker bit is $0$ from greatest to least RRPV (BARD-C). Ties for RRPV values are broken arbitrarily. If the victim is dirty but its BLP-Tracker bit is $1$, then BARD searches for another dirty line with a BLP-Tracker bit of $0$ from greatest to least RRPV (BARD-E). 

Figure~\ref{fig:srrip_results} shows the performance of BARD under LRU, SRRIP~\cite{jaleel2010srrip}, and SHiP~\cite{wu2011ship}; the latter two replacement policies use RRIP. For each BARD design, we show the normalized performance with respect to a system that uses the same replacement policy. On average, BARD provides improvements of 4.3\%, 5.0\%, and 4.9\% when using LRU, SRRIP, and SHiP, respectively. So, BARD remains effective even for modern replacement policies that use RRIP~\cite{wu2011ship, shah2022mockingjay}. 

\subsection{Impact of a Larger Core Count}
\label{sec:16core_study}
Table~\ref{tab:bard_core_count_study} shows the gmean and maximum speedups achieved by BARD for the 8-core (baseline) and 16-core configurations. Note that the 16-core configuration uses a 32MB LLC and 2 DDR5 channels. For the 8-core system, BARD's speedups are 4.3\% on average and up-to 8.5\%. For the 16-core system, BARD's speedups are 5.5\% on average and up-to 11.5\%. BARD is more effective for systems with more cores as the additional cores in these systems generate more memory-related traffic (both read and write). We expect BARD perform as well, or better, for even larger systems.

\begin{table}[!htb]
    \begin{center}
    \vspace{-0.1in}
    \caption{BARD Speedup on 8 and 16 Core Systems}
    \label{tab:bard_core_count_study}
    \begin{tabular}{|c||c|c|}
        \hline
        \textbf{Core Count} & \textbf{Gmean} & \textbf{Max}  \\
        \hline
        \hline
        8 & 4.2\% & 8.8\% \\
        16 & 5.1\% & 11.1\% \\
        \hline
    \end{tabular}
    \vspace{-0.2in}
    \end{center}
\end{table}

\subsection{Impact of Write Queue Size}
    Figure~\ref{fig:write_buffer_sensitivity} compares the performance of the baseline
    and BARD for different write queue sizes. All results are normalized to a baseline
    with a 48-entry write queue (the baseline configuration). As shown, the baseline
    has speedups of -6.2\%, 0.0\%, 3.3\%, 8.1\%, and 10.7\%, respectively. BARD
    has speedups of 0.4\%, 4.3\%, 7.0\%, 10.0\%, and 11.7\%, respectively.
    BARD provides comparable speedups to increasing the capacity of the write queue at
    much lower hardware overhead. For example, increasing the write queue capacity from
    48 entries to 64 entries for both sub-channels within a channel consumes about 4.4KB
    and improves performance by 3.3\%. In contrast, using BARD only incurs an overhead
    of 8B per LLC slice and improves performance by 4.3\%.
    
    \begin{figure}
        \centering
        \includegraphics[width=\columnwidth]{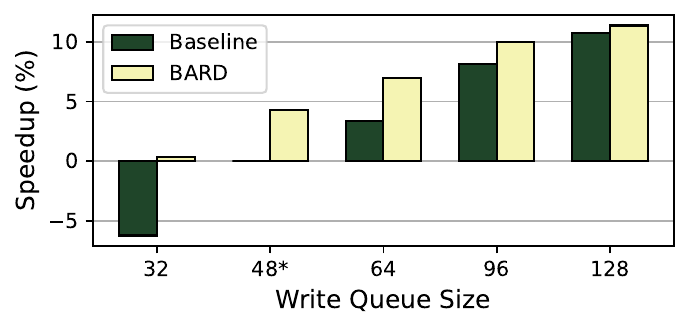}
        \caption{Speedups of baseline and BARD for 32, 48, 64, 96, and 128 entry write
                queues.
                The baseline has speedups of -6.2\%, 0.0\%, 3.3\%, 8.1\%, and 10.7\%,
                respectively.
                BARD provides speedups of 0.4\%, 4.3\%, 7.0\%, 10.0\%, and 11.7\%,
                respectively.}
        \label{fig:write_buffer_sensitivity}
    \end{figure}

\subsection{Synchronization Overheads of BARD}
\label{sec:bard_sync_overheads}

Whenever the LLC performs a writeback, BARD may broadcast the corresponding line's bank address so each BLP-Tracker has up-to-date information about which DRAM banks have pending writes\footnote{While BARD need not send a broadcast if the bank's BLP-Tracker bit is already $1$, we observe that almost all writebacks require a broadcast. So, for simplicity, we assume each writeback requires a broadcast.}. In this section, we analyze how much bandwidth BARD consumes when performing these broadcasts. We perform our analysis for a high-end server processor with 128 cores and 8 memory channels. As our evaluations consider 8 cores, we assume that the 128-core processor has $16\times$ more write traffic than the system evaluated in our paper.

First, note that on every writeback, the LLC must transmit 70B to the appropriate memory channel. Six of these bytes are the physical address of the writeback, and 64B is the data of the writeback. These overheads are incurred in both the baseline and in BARD. The only additional bandwidth consumed by BARD is used to broadcast the bank address of the writeback to each LLC slice. In a processor with 8 DDR5 channels, this is a 9-bit address as there are 512 total DRAM banks across all 8 channels. Hence, after each writeback, BARD sends slightly more than 71B of data.

Table~\ref{tab:bard_bw} shows the bandwidth consumption of BARD, divided into bandwidth required for issuing writebacks and bandwidth required for synchronizing the BLP-Trackers. Performing writebacks consumes 153.9 GB/s of bandwidth on average, and 281.1 GB/s in the worst case -- note that this is also required in the baseline. Synchronizing the BLP-Trackers requires 2.5 GB/s extra bandwidth on average, and 4.5 GB/s in the worst case. \textit{This is only a 1.6\% increase in total bandwidth consumption}. For comparison, a NoC for a 128-core system must already have multiple TB/s of bandwidth to support the frequent coherence updates caused by each core's private caches. So, BARD's synchronization overheads are negligible.

\begin{table}[!htb]
    \begin{center}
    \caption{BARD Bandwidth Overheads}
    \label{tab:bard_bw}
    \begin{tabular}{|c||c|c|c|}
        \hline
        \textbf{Purpose} & \textbf{Packet Size} & \textbf{Mean (GB/s)} & \textbf{Max (GB/s)} \\
        \hline
        \hline
        Writeback & 70B = 560b & 153.9 & 281.3 \\
        Synchronization & 9b & 2.5 & 4.5 \\
        \hline
    \end{tabular}
    \end{center}
\end{table}

\vspace{0.1in}
\noindent
\textbf{When to send the bank address:} Sending a writeback's bank address \textit{after} eviction is too late, as other LLC slices may also have performed an eviction whose writeback maps to the same bank. A na\"{i}ve, yet ineffective, solution would be serializing the evictions in each LLC slice: this degrades performance. Ideally, we want each LLC slice to be aware of other slices' decisions on time without degrading performance.

Note that eviction happens in two stages: victim selection and the corresponding fill. Victim selection begins after a miss, whereas a fill begins after the LLC receives data from DRAM. As DRAM is many times slower than the CPU, there are hundreds of cycles between the end of victim selection and the beginning of a fill. Our insight is that a writeback's bank address should be transmitted after victim selection, instead of at the end of the eviction. Now, if two slices initially attempt a writeback to the same DRAM bank, one slice will complete victim selection first and broadcast the bank it decides to issue a writeback to. The second slice can now update its BLP-Tracker and restart victim selection. In the unlikely scenario where an LLC slice cannot select a victim before the beginning of the fill, it can default to evicting the LRU victim.

\vspace{0.1in}
\noindent
\textbf{Multi-Node Synchronization:} We further discuss how BARD scales to a multi-node system. Unfortunately, synchronizing the BLP-Trackers in completely different processors is slow and impractical. However, note that a remote writeback (one node issues a writeback to a memory channel connected to another node) must first check the LLC for a hit. Suppose the writing node is $A$ and the receiving node is $B$. If the writeback from $A$ hits in $B$'s LLC, then the data is updated. If the writeback misses the LLC, then the writeback is installed anyway (some line is replaced). Either way, the writeback from $A$ will be installed in $B$'s LLC. Thus, even if the BLP-Trackers on $A$ are not synchronized with the BLP-Trackers on $B$, the BLP-Trackers on $B$ will ensure that the LLC's writeback stream is optimized for bank-level parallelism. \textit{Hence, the BLP-Trackers on different nodes need not be synchronized.}

\subsection{Analyzing the BLP-Tracker's Accuracy}
\label{sec:blp_tracker_accuracy}

While the BLP-Tracker is low-cost, it may not be accurate as it does not communicate with the processor's memory controllers. Ideally, when BARD overrides the underlying replacement policy (BARD-E) or cleanses a line, the new writeback should not map to a bank with a write pending in one of the memory controller's write queues. Our studies suggest that on average, BARD makes incorrect decisions 30.3\% of the time, suggesting that BARD's mechanism can be improved. However, while BARD's current implementation is imprecise, it provides large improvements in bank parallelism (22.4 to 28.9) and a substantial speedup (4.3\%) given its minor overheads (8B per LLC slice). We hope that future work will improve the accuracy of the BLP-Tracker mechanism.

\subsection{Impact on DRAM Power Consumption}
\label{sec:bard_power}

Table~\ref{tab:bard_power} compares the mean power consumption, energy consumption, and energy-delay product (EDP) for BARD and the Virtual Write Queue (VWQ) normalized to the baseline. For all metrics, lower is better. While BARD consumes slightly more energy than VWQ, it has a lower EDP (0.970 versus 0.995) since it provides a substantial speedup (4.3\%), whereas VWQ provides a slowdown (-0.3\%).

\begin{table}[!htb]
    \begin{center}
    \caption{Power, Energy, and EDP norm. to Baseline}
    \label{tab:bard_power}
    \begin{tabular}{|c||c|c|c|}
        \hline
        \textbf{System} & \textbf{Power} & \textbf{Energy} & \textbf{EDP} \\
        \hline
        \hline
        BARD & 1.06 & 1.015 & 0.970 \\
        VWQ & 0.989 & 0.993 & 0.995 \\
        \hline
    \end{tabular}
    \end{center}
\end{table}

\subsection{Impact on LLC Misses and Writebacks}
\label{sec:bard_misses_and_writebacks}
    Table~\ref{tab:bard_misses_and_writebacks} shows the relative MPKI and WPKI of BARD to the baseline.
    For misses, BARD reduces MPKI by 0.6\% on average, and in the worst case increases MPKI by 1.4\%. 
    For writebacks, BARD increases WPKI 2.7\% on average and up-to 8.3\%.
    BARD's reduction in MPKI is likely due to the ineffectiveness of LRU on several workloads.
    Hence, the results for MPKI suggest that BARD does not significantly affect cache miss rate.
    In contrast, BARD does increase the number of writebacks significantly.
    We observe that even on workloads where BARD significantly increases writebacks, the additional writebacks do not cause slowdown as BARD significantly improves the BLP for such workloads. For example, on \texttt{cam4} where BARD causes the most additional writebacks (8.5\%), BARD increases the bank-level parallelism from 22.0 to 29.5. As a result, despite the additional writes, BARD spends less time writing to DRAM than the baseline.

    \begin{table}[!htb]
        \begin{center}
        \caption{Misses and Writebacks relative to Baseline}
        \label{tab:bard_misses_and_writebacks}
        \begin{tabular}{|c||c|c|}
            \hline
            \textbf{Metric} & \textbf{Mean} & \textbf{Max}\\
            \hline
            \hline
            Misses & 0.0\% & 1.3\% \\
            Writebacks & 2.7\% & 8.5\% \\
            \hline
        \end{tabular}
        \end{center}
    \end{table}

\section{Related Work}
\label{sec:related}

The two works most closely related to our proposal are: {\bf{Eager Writeback (EW)}}~\cite{lee2000eagerwriteback} and {\bf{Virtual Write Queue (VWQ)}}~\cite{stuecheli2010virtualwritequeue}.  We compare BARD both qualitatively and quantitatively with them in Section~\ref{sec:vwq}. Our study shows that the assumptions for both EW and VWQ are ineffective in DDR5, and they both provide slowdowns of 0.5\% on average. In contrast, BARD exploits BLP for write requests, and this is a more effective way to mitigate write-related stalls in DDR5.  

\ignore{
In this section, we discuss prior research related to the general trends of mitigating DRAM write overheads and/or mitigating bank or bankgroup related overheads.

\subsection{Expanding the Write Buffer}
    As discussed previously, the small capacity of the write buffer prevents effectively leveraging DRAM resources, namely bank-level parallelism. We discuss prior works that aim to expand the effective capacity of the write buffer by using the LLC.
}    
 
\vspace{0.05 in}   

\textbf{DRAM-Aware LLC Writeback}~\cite{lee2010dramawarewriteback} attempts to increase row buffer hit rate by monitoring accesses to dirty lines that mapped to currently open DRAM rows.
\textbf{Last Level Write Predictor (LLWP)}~\cite{wang2012lastlevelwritepredictor} identifies last-write to cache lines with a dead block predictor~\cite{khan2010samplingdeadblockpredictor}. The dirty lines that are predicted to be dead can be written when convenient. The LLWP requires significant storage overheads (16-bit signatures of PC for each cache line in the private caches), and LLWP does not specifically optimize BLP for write operations. \textbf{Dirty Block Index (DBI)}~\cite{seshadri2014dirtyblockindex} removes dirty line information from the tag-store of the LLC and instead tracks dirty lines by DRAM row in a separate, associative structure. DBI tries to do aggressive writeback (similar to VWQ) whereby dirty lines to rows with pending writes are proactively written back (to increase row-buffer hits). Unfortunately, in DDR5, focusing on row-buffer hits for write operations is not as effective as writing to the same bankgroup can only occur once every 20ns (6x higher latency than consecutive writes to different bankgroups). BARD exploits BLP to reduce write latency. BARD is also less costly than prior work (8B of SRAM per channel per LLC slice vs multiple KB).




\textbf{Bank-Level Parallelism Aware Prefetch Issue (BAPI)}~\cite{lee2000eagerwriteback} is a prefetching strategy that aims to maximize bank-level parallelism to improve prefetch timeliness. Prefetches that poorly exploit bank-level parallelism fail to complete quickly, thus failing to hide memory access latency. The \textbf{Duplicon Cache}~\cite{lin2018duplicon} duplicates data within DRAM to improve bank-level parallelism. Both BAPI and Duplicon Cache improve bank-level parallelism for prefetches and demand accesses, respectively.
In contrast, BARD seeks to improve bank-level parallelism for write requests.
    

\vspace{0.05 in}
Several works, such as \textbf{SHiP}~\cite{wu2011ship}, \textbf{HawkEye}~\cite{jain2016hawkeye}, and \textbf{Mockingjay}~\cite{shah2022mockingjay}, optimize the RRIP-based policy to reduce cache misses. These policies tune the insertion decision for RRIP (install with $\mathrm{RRPV}=3$ or $\mathrm{RRPV} < 3$) based on a prediction mechanism. Our implementation for BARD with RRIP is still applicable to such policies.  Unlike standard replacement policies, whose goal is to reduce cache misses, {\em cost-aware replacement policies} aim to minimize the average access time by evicting {\em easy-to-fetch} lines over {\em hard-to-fetch} lines~\cite{jeong2003costawarecachereplacement}. \textbf{MLP-Aware Cache Replacement}~\cite{qureshi2006mlpawarecachereplacement} seeks to avoid isolated misses and thus improve memory-level parallelism. Similarly, BARD is also a cost-aware replacement policy, but unlike prior work, BARD aims to optimize write bank-level parallelism.

\section{Conclusion}

In this paper, we study the problem of stalls due to DRAM writes for DDR5 systems. DDR5 worsens the DRAM timing parameters and causes variable write-to-write delays, depending on consecutive writes, depending on whether they go to different bankgroups (1x), same bankgroup (6x), or same bank with a row-buffer conflict (24x).  We show that the memory system spends 33\% of the time doing DRAM writes. The key insight of our work is to perform {\em Bank-Aware Replacement Decisions (BARD)} and modify the replacement policy to provide write requests that would incur lower write latency. BARD exploits the fact that DDR5 greatly increased the number of banks (32 per sub-channel, 64 per channel) thus providing abundant bank-level parallelism (BLP). BARD improves the BLP for write requests by steering write traffic to banks that have no pending writes.  BARD improves the BLP for writes from 22.1 to 28.8 (out of a maximum of 32) and provides an average performance improvement of 4.3\%.  We show that BARD is highly effective for DDR5 systems compared to prior proactive writeback schemes. BARD requires only 8 bytes per channel per LLC slice.

\section*{Acknowledgements}

We thank the reviewers of MICRO 2025 and HPCA 2026 for their valuable feedback. We also thank Hritvik Taneja (GT), Anish Saxena (GT), and Jae Hyung Ju (GT) for helpful discussions.  This work was supported by NSF grant 233304.

%
%
%
%
%


\section*{Artifact Appendix}

\subsection{Abstract}

Our artifacts are the modified Champsim~\cite{gober2022champsim} simulator and the evaluated traces.
Our modifications to Champsim include (1)~modifications to the cache implementation to support Virtual Write Queue and Eager Writeback, and (2)~a new DRAM model that models DDR5 timing faithfully.

\subsection{Artifact check-list (meta-information)}

{\small
\begin{itemize}
  \item {\bf Program:} {various executables are generated by Champsim}
  \item {\bf Compilation:} \texttt{gcc}, at least version 12
  \item {\bf Data set:} Champsim Traces for SPEC2017, LIGRA, STREAM, and Google server traces
  \item {\bf Hardware:} Compute cluster required
  \item {\bf Execution:} Python scripts automate experiments
  \item {\bf How much disk space required (approximately)?:} 50GB is a safe bet.
  \item {\bf How much time is needed to prepare workflow (approximately)?:} An hour, which is mostly spent doing compilation.
  \item {\bf How much time is needed to complete experiments (approximately)?:} At most 3 days
  \item {\bf Publicly available?:} Yes, codebase is on \href{https://github.com/suhaskvittal/champsim-svittal}{Github}. 
  \item {\bf Code licenses (if publicly available)?:} MIT
  \item {\bf Data licenses (if publicly available)?:} MIT
  \item {\bf Archived (provide DOI)?: } Traces are archived on Zenodo: https://doi.org/10.5281/zenodo.17862160.
\end{itemize}
}

\subsection{Description}

\subsubsection{How to access}

Codebase is available on \href{https://github.com/suhaskvittal/champsim-svittal}{Github}. 
Use the \texttt{bard\_hpca2026} branch (run \texttt{git checkout -b bard\_hpca2026}) as we may modify the \texttt{master} branch in future work.
Traces are available on \href{https://doi.org/10.5281/zenodo.17862160}{Zenodo}.

\subsubsection{Hardware dependencies}
    Running experiments will require a compute cluster or a set of servers.
    If hundreds of cores cannot be allocated at once, experiments may have to be batched manually.
    
\subsubsection{Software dependencies}
    Our code compiles with \texttt{gcc-12} through \texttt{gcc-14}. It has not been tested with \texttt{clang}. Champsim uses vcpkg as a dependency. Please initialize the vcpkg submodule as shown in the README.

    Our code uses python scripts to generate statistics and create plots. Our dependencies are numpy, scipy, pandas, and matplotlib.

\subsubsection{Data sets}
    Our evaluated traces are provided on Zenodo. For this artifact, do not modify the structure of these traces as all scripts rely on this structure.

\subsection{Installation}

\subsubsection{Initial Steps}
    Install vcpkg as shown in the Champsim README.
    Once this is done, create all configuration files by running the \texttt{generate\_configs.py} script (run from the base directory).
    This should populate the json folder with the appropriate configurations.

\subsubsection{Building Executables}
    Once the configuration files are built (should take less than a second),
    run the \texttt{build\_all.py} script.
    This will compile all Champsim executables used in this artifact using 8 cores.
    This will likely take an hour.

\subsubsection{Running Experiments}
    First, run the \texttt{all\_runs.sh} script (again, from the base directory).
    This creates a \texttt{commands.out} file that contains a list of
    all commands necessary to run the appropriate experiments.
    If desired, this file can used by GNU parallel to run experiments
    (i.e., \texttt{parallel -a commands.out}).
    However, on most computing clusters, this is not possible.
    Our cluster uses SLURM, and we have a utility script called
    \texttt{run\_command\_list\_on\_pace.py} that runs all commands
    in \texttt{commands.out} on our cluster.
    This script configures the walltimes of each command depending
    on the number of cores detected.
    8-core configurations will take about 24 hours in the worst case.
    16-core configurations will take about 3 days in the worst case.

    To run a partial set of experiments, modify \texttt{all\_runs.sh}
    by commenting out given lines.

\subsection{Experiment workflow}
    \subsubsection{Generating Statistics Files}
        The Champsim simulator returns many statistics.
        We provide a script called \texttt{collect\_stats.py} that
        aggregates the statistics relevant to our project in CSV
        files.
        This script handles computing weighted speedup and other
        statistics of interest that are not directly reported
        by the simulator.
        These files are used when reporting results in the paper.
    
    \subsubsection{Figures}
        All figures can be plotted using the notebook \texttt{bard\_plots.ipynb} in the scripts folder.
        The notebook specifically generates plots for Figures
        10, 11, 14, and 15.

    \subsubsection{Tables}
        The \texttt{bard\_report\_stats.py} scripts reports
        geomean and maximum statistics for all plots as well
        as other statistics found in each table.
        Data reported in Tables VI, VII, and VIII, and IX
        are reported by this script (along with any geomean/max
        numbers quoted in Figure captions).

\subsection{Evaluation and expected results}
    Generated plots and data should roughly match what is reported in the main text, with some possible variance due to randomness.



\subsection{Methodology}

Submission, reviewing and badging methodology:

\begin{itemize}
  \item \url{https://www.acm.org/publications/policies/artifact-review-and-badging-current}
  \item \url{https://cTuning.org/ae}
\end{itemize}


\balance
\bibliographystyle{IEEEtranS}
\bibliography{refs}

\end{document}